\documentstyle[preprint,aps,tables,tighten,psfig,epsf]{revtex}
\begin{document}
\draft
\preprint{\centerline{Submit to Physical Review D}}
\title{Extracting the  Proton $\bar{u}$ content\\ 
from  pp $\rightarrow$ Direct Photon plus Jet Cross Sections.\\}
\author{R.~G.~Badalian}
\address{Laboratory of High Energy\\
Joint Institute for Nuclear Research\\
141980 Dubna, Moscow Region, Russia}
\author{S.~Heppelmann}
\address{Department of Physics\\
Pennsylvania State University\\
University Park, Pennsylvania 16802, USA}
\date{\today}
\maketitle
\begin{abstract}
An analysis procedure is proposed to measure
the antiquark distributions 
in the proton over the region $0.01 < x < 0.1$.  The procedure involves 
the measurement of high $p_t$ asymmetric direct photon and jet final 
states in $pp$ interactions. 
This measurement can be made at the $RHIC$ collider running in $pp$
mode at an energy of $\sqrt{s}=500~GeV/c$. 
This analysis identifies a region
of phase space  where
the contribution from
quark--antiquark annihilation 
uncharacteristically approaches the magnitude of the contribution from 
the leading process,  quark--gluon 
Compton scattering. 
The forward-backward angular asymmetry in the parton center of mass 
is sensitive to the antiquark content of
the proton and the $\bar{u}$ parton density function can be extracted. 
\end{abstract}
\newpage

\section{Introduction}
\label{sec:intro}
The distribution functions of quarks and gluons in a proton 
play a crucial role in the self consistent description 
of all hard inclusive proton interactions. 
The cross sections in both lepton--proton and proton--proton high--$p_t$
interactions are derived from the fundamental perturbative $QCD$ 
parton interaction cross sections, folded with the parton distribution 
functions and integrated over available phase space. 
The main sources of experimental input for the parton distribution 
functions (or structure functions) include
deep inelastic lepton--nucleon scattering and hard 
hadronic inclusive interactions
with final states including  high mass leptons 
pairs, high--$p_t$ direct photons, 
high--$p_t$ jets, $W$, $Z$ or heavy flavors.

Knowledge of the quark and gluon distributions of the proton 
has greatly increased in recent years. 
The first input has come from the high 
quality data on deep--inelastic structure functions
in lepton--nucleon scattering \cite{h1,zeus}. 
These experiments are generally most sensitive to
valence quark distributions. They probe
electromagnetically and are only directly sensitive to charged partons 
rather than gluons.
The deep--inelastic scattering experiments have been less
conclusive in the determination of the antiquark distributions.
Analyses of data to extract the antiquark
distributions usually involve various assumptions \cite{duke}. 

Progress in determining the sea quark 
and gluon distributions in the proton have concentrated on fits to
a variety of types of data.
The $CTEQ$ Collaboration \cite{cteq95,cteq93,morfin} 
and $MRS$ group \cite{martin} have made  global fits to  available 
experimental data. But, the determination of the antiquark distributions with 
hadronic interactions using fit procedures has been difficult 
because most high--$p_t$ hadronic reactions are either: 
dominated by contributions from
valence partons and gluons; 
involve too many contributing subprocesses 
to analyze unambiguously; 
or have a very small cross section in the high--$p_t$ region. 
Furthermore, much of the recent high energy collider data 
involves $\bar{p} p$ 
interactions where the most probable interactions tend to
involve gluons and valence quarks rather than non--valence quarks.

It is unclear how well the antiquark distributions 
have been determined in the fits by
the $CTEQ$ and $MRS$ groups.
Even when considering similar data sets, these groups find that
their fits differ in the low $x$ region \cite{cteq95}. 
For example, at $Q^2 = 25 \, GeV^2$ the $\bar{u}(x,Q^2)$ distribution from 
$MRS$ at $x \simeq (0.01 - 0.02)$ is higher by $(10 - 15)\%$ than the same 
distribution of the $CTEQ$ Collaboration. For the $d$ antiquark 
$\bar{d}(x,Q^2)$ distribution, 
this difference increases up to $(15 - 20)\%$ at same region of $x$ and 
$Q^2$. It has been difficult to pin down the errors associated with
these fits.

To verify and improve the results of these or other fits, 
measurements which are more directly
dependent upon and more sensitive to the antiquark distributions are needed.
In this paper, one such experiment and analysis procedure is proposed.
A kinematic region is defined for  high--$p_t$ direct photon and jet 
production from $pp$ interactions 
in which the $u \bar{u}$ annihilation process is uncharacteristically 
important. The region involves parton pairs 
with asymmetric momentum fractions and
the analysis concentrates on the angular distribution of the parton--parton 
interaction.

In general, the dominant subprocess for high--$p_t$ direct photon and 
jet production is the $u$ quark--gluon Compton process. Since the
direct photon  cross section is 
proportional to the square of the quark or antiquark charge, 
the $d$ and $s$ quark 
contributions are suppressed by a factor of four relative to that of the
$u$ quarks. 
Thus,  $u + G \rightarrow u + \gamma$  is the primary subprocess and
the next most 
important process is  $u$ and $\bar{u}$ quark annihilation, 
$u + \bar{u} \rightarrow G + \gamma$. 
The $u$ quark contribution to direct photon and jet production
is enhanced over the  $d$ or 
$s$ quark contributions  both because of this factor of four 
and because of the reduced number of $d$ or $s$ quarks in the proton
relative to $u$ quarks. 

While the annihilation parton cross  section is comparable in magnitude 
to that of the Compton process, the ratio of the distribution 
functions $\bar{u}(x)$ to $G(x)$ is typically 1:10 in the region of 
$x \simeq 0.01 - 0.1$. Generally, the $u$ and $\bar{u}$ annihilation 
contribution to the direct photon cross section is only about $10 \%$. 
These two types of  parton 
cross sections do, however, have a different angular distribution.

In the parton model, the direct photon--jet cross section is the
sum over subprocess terms, each term is the product of
a parton level cross section and the two
associated parton density functions. 
The lowest order partonic cross sections for annihilation and Compton 
interactions are \cite{halzen,owens}: 
\begin{equation}
q + \bar{q} \rightarrow G + \gamma  \hskip 0.3 in \Longrightarrow 
\hskip 0.3 in {d\sigma \over {d\hat{t}}} \sim 
{8 \over {3 \, \hat{s}^2}} \, {\left({{\hat{t}^2 + \hat{u}^2} 
\over {\hat{t}\hat{u}}}\right)} 
\label{eq:number1}
\end{equation}

\begin{equation}
q + G \rightarrow q + \gamma \hskip 0.3 in \Longrightarrow \hskip 0.3 in 
{d\sigma \over {d\hat{t}}} \sim - \, 
{1 \over {\hat{s}^2}} \, {\left({{\hat{s}^2 + \hat{u}^2} 
\over {\hat{s}\hat{u}}}\right)}, 
\label{eq:number2}
\end{equation}
where $\hat{s}$, $\hat{t}$ and $\hat{u}$ are the Mandelstam variables for 
the parton scattering subprocess. 

These two cross sections have very different angular distributions. 
The Compton cross section becomes very large for scattering 
angles corresponding 
to the $u$ quark scattering backward relative to the incident $u$ quark
direction, sending the photon forward.
The annihilation cross section becomes large for the photon 
emerging in the forward or backward direction. 
The idea of this analysis
procedure is to focus on  $\gamma + jet$ production in the region 
of the parton center of mass angular distribution, where the annihilation 
process is enhanced because of the small $\hat{t}$ singularity in the
low order scattering graph.

The variables $x_1$ and $x_2$ are defined
to be the momentum fractions 
for  the  ``first'' and ``second'' protons 
respectively, and $\vartheta_{\gamma}$ as the angle between the photon and 
the ``first'' proton momentum direction in the center of mass of a partonic 
hard scattering subprocess. Exact definition of ``first'' and ``second'' 
protons will be explained in section \ref{sec:pythia}. 
From observable rapidities $\eta_{\gamma}$, $\eta_{jet}$, the
transverse momentum $p_t$ and the $pp$ center of mass energy $\sqrt{s}$, 
the quantities 
$x_1$, $x_2$ and $\vartheta_{\gamma}$ are defined as: 

\begin{equation}
x_1 = {p_t \over {\sqrt{s}} }{ (e^{ \eta_{\gamma}} + e^{\eta_{jet}})} 
\label{eq:number3}
\end{equation}

\begin{equation}
x_2 = {p_t \over {\sqrt{s}}}{ (e^{ -\eta_{\gamma}} + e^{-\eta_{jet}})} 
\label{eq:number4}
\end{equation}

\begin{equation}
\tan \left( \vartheta_{\gamma} \over 2 \right) = 
\exp \left( {{\eta_{jet} - \eta_{\gamma}} \over 2} \right) .
\label{eq:number5}
\end{equation}

The forward--backward asymmetry $A(x_1,x_2)$ is then defined as the ratio of the 
forward to backward cross section: 

\begin{equation}
A(x_1, x_2) = {{\sigma (\cos (\vartheta_{\gamma}) > 0.7)} \over 
               {\sigma (\cos (\vartheta_{\gamma}) < -0.7)}}
\label{eq:number6}
\end{equation}
The choice of $\cos (\vartheta) = \pm 0.7$ as the boundary defining
the forward
or backward regions, is somewhat  arbitrary.

For the two main subprocesses, annihilation and Compton scattering,
the forward--backward 
asymmetry will have a different magnitude. 
The pair annihilation subprocess 
$A_{q\bar{q}}(x_1, x_2) = 1$, and the Compton scattering case 
$A_{qG}(x_1, x_2) \gg 1$. Using this fundamental difference 
between $A_{q\bar{q}}$ and $A_{qG}$, the relative strengths of the 
$\bar{q}(x)$ and $G(x)$  distributions in the proton can be determined.

\section{Analysis of Angular Asymmetry}
\label{sec:analasym}
This analysis  considers the direct photon and jet production 
cross section from $pp$ collisions in a
small kinematic region defined by 
$x_1 \pm \Delta x$ and $x_2 \pm \Delta x$ where $x_1 > x_2$. 
The asymmetric condition 
$x_1 > x_2$ is important, enabling the observation 
of the forward--backward asymmetry of the direct photon 
and jet angular distribution.
The proton antiquark density  $\bar{u}(x_2)$ will be extracted from
the direct photon forward--backward asymmetry in $pp$ collisions as 
introduced in 
equation \ref{eq:number6}. 

\begin{figure}[tb]
        \begin{center}
        \leavevmode 
        \epsfxsize=0.8\hsize
        \epsfbox{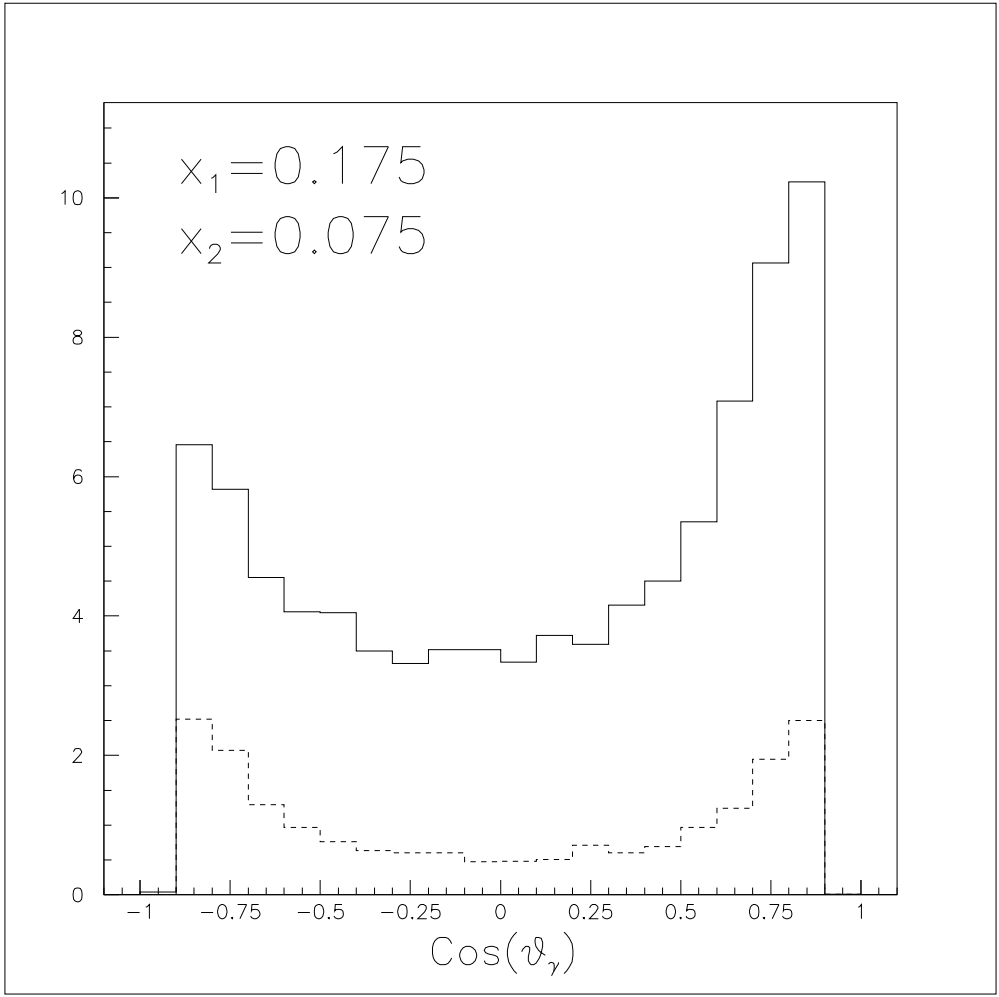}
\end{center}                         
\caption{The $\cos(\vartheta_{\gamma})$ distributions at $x_1 = 0.175$ and 
$x_2 = 0.075$ from the reaction $pp \rightarrow \gamma + jet$ at $pp$ 
center of mass energy $500~GeV$ and $p_t > 10 ~GeV$ in the acceptances of 
$-1 < \eta_{\gamma} < 2$ and $-1 < \eta_{jet} < 2$ for photon and 
jet respectively. 
Solid curve -- Compton scattering subprocess, dashed curve -- pair 
annihilation subprocess. The numbers in the vertical axis are cross sections in picobarns. 
}
\label{fig:fig3}
\end{figure}

For illustration, consider the region
$x_1 = 0.175 \pm 0.025$ and $x_2 = 0.075 \pm 0.025$.  
The  $\cos(\vartheta_{\gamma})$ distributions for the two main 
subprosses from a $PYTHIA$ simulation of the direct photon and jet 
production in $pp$ interactions \cite{sjostrand} are plotted in figure 
\ref{fig:fig3}. The $pp$ center of mass energy is $\sqrt{s} = 500~GeV$ 
and the events are selected to be in the kinematic region defined by 
$p_t > 10~GeV$,  $-1 < \eta_{\gamma} < 2$ 
and  $-1 <  \eta_{jet} < 2$ for both the 
jet and the photon.  In the region $\cos(\vartheta_{\gamma}) > 0.7$ the 
contribution from the Compton scattering subprocess $qG$ is more 
then five times greater than the contribution from the pair annihilation 
subprocess $q \bar{q}$. In contrast, for $\cos(\vartheta_{\gamma}) < -0.7$ 
the  contribution from the $qG$--subprocess is only about twice the 
$q \bar{q}$--subprocess contribution. 

\begin{figure}[tb]
        \begin{center}
        \leavevmode 
        \epsfxsize=0.8\hsize
        \epsfbox{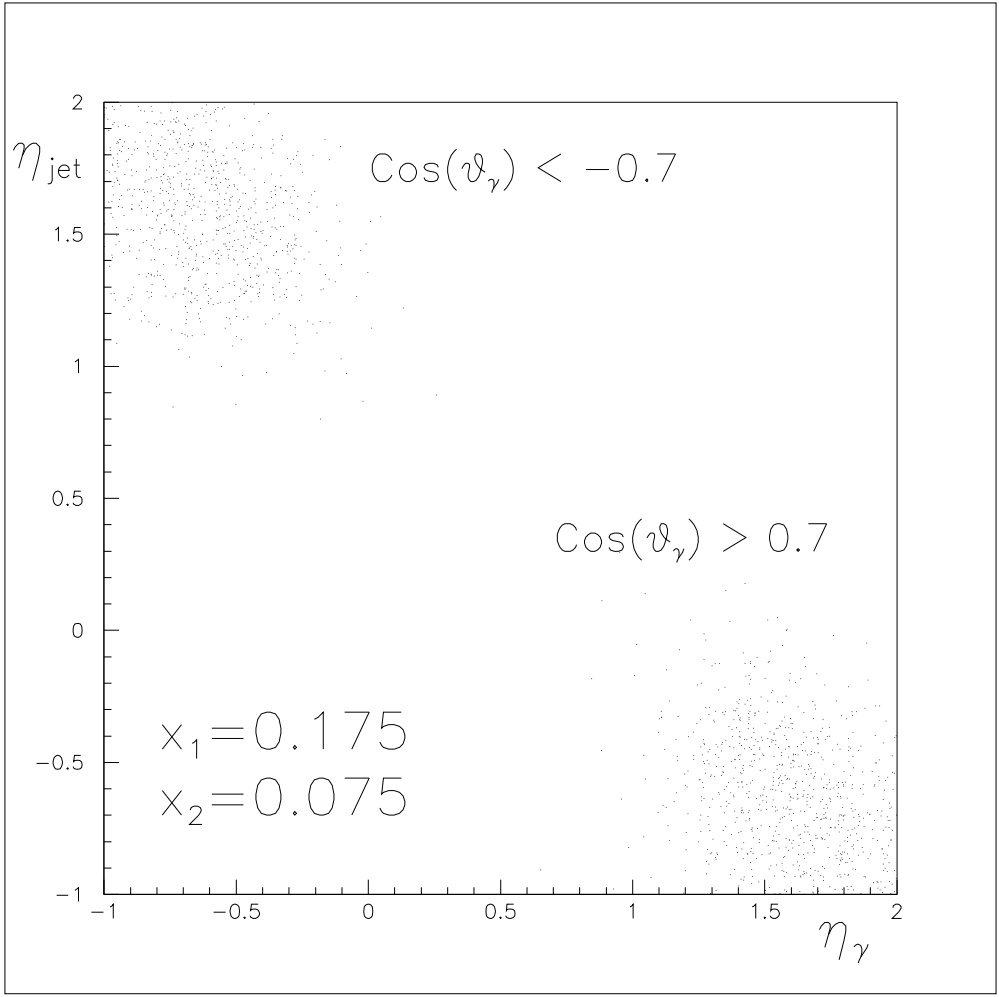}
\end{center}                         
\caption{Events distribution on a $(\eta_{\gamma} \times \eta_{jet})$ plot 
from the reaction $pp \rightarrow \gamma + jet$ at $pp$ center of mass 
energy $500~GeV$ and $p_t > 10 ~GeV$ in the acceptances of 
$-1 < \eta_{\gamma} < 2$ and $-1 < \eta_{jet} < 2$ for photon and 
jet respectively. 
The events are from $x_1 = 0.175 \pm 0.025, \, x_2 = 0.075 \pm 0.025$. 
In the top part of the figure events are from 
$\cos(\vartheta_{\gamma}) < -0.7$ and in the bottom part of the 
figure from $\cos(\vartheta_{\gamma}) > 0.7$.
}
\label{fig:fig4}
\end{figure}

The two dimensional plots of the event distributions under the conditions 
$\cos(\vartheta_{\gamma}) > 0.7$ and $\cos(\vartheta_{\gamma}) < -0.7$ are 
plotted in figure \ref{fig:fig4} for the same region of 
$x_1 = 0.175 \pm 0.025$ and $x_2 = 0.075 \pm 0.025$.  
From the results presented in figure \ref{fig:fig4}, we see that 
the $\cos(\vartheta_{\gamma}) > 0.7$ cut selects large photon rapidity
and small jet rapidity  while the $\cos(\vartheta_{\gamma}) < -0.7$ cut
selects the reverse ordering.
Refining the definition of the asymmetry variable to include experimental 
acceptance, $A(x_1, x_2)$ is to be the  accepted cross section (or events 
number) for $\cos(\vartheta_{\gamma}) > 0.7$ divided by the accepted cross 
section  $\cos(\vartheta_{\gamma}) < -0.7$ for a particular $(x_1, x_2)$ bin. 

In terms of the parton distribution functions  $i(x)$ for parton $i$ 
and the subprocess differential
cross sections ${d\sigma_{i,j}} / {d\cos(\vartheta_{\gamma})}$ 
for the process $i+j\rightarrow jet+\gamma$ , the asymmetry is 

\begin{equation}
A(x_1, x_2) \, = \,
\frac
{\sum \limits_{i,j} i(x_1) \, j(x_2) \, \sigma^{+}_{i,j}}
{\sum \limits_{i,j} i(x_1) \, j(x_2) \, \sigma^{-}_{i,j}}\,
= \,
\frac
{\sum \limits_{i,j} N^+_{i,j}(x_1,x_2)}
{\sum \limits_{i,j} N^-_{i,j}(x_1,x_2)}\, ,
\label{eq:number7}
\end{equation}
where $N^{\pm}_{i,j}(x_1, x_2) = i(x_1) \, j(x_2) \, 
\sigma^{\pm}_{i,j}$ 
and $i$, $j$ = $u,\bar{u},d,\bar{d},s,\bar{s},c,\bar{c}$ and $G$. 
The cross--sections $\sigma^{\pm}_{i,j}$ are defined as: 

\begin{equation}
\sigma^{+}_{i,j} \, = 
\int \limits_{\cos(\vartheta_{\gamma}) > 0.7} 
\frac{d\sigma_{i,j}}{d\cos(\vartheta_{\gamma})}
d\cos(\vartheta_{\gamma}) 
\label{eq:number8}
\end{equation}

\begin{equation}
\sigma^{-}_{i,j} \, = 
\int \limits_{\cos(\vartheta_{\gamma}) < -0.7} 
\frac{d\sigma_{i,j}}{d\cos(\vartheta_{\gamma})}
d\cos(\vartheta_{\gamma}). 
\label{eq:number9}
\end{equation}

In an actual experiment, the divergence at the $\pm 1$ limits of integration 
will be removed  by acceptance and experimental cuts.

It is possible to estimate the asymmetry in the limit of
full acceptance, with cross sections dominated by the small $t$ and 
small $u$ singularities of equations \ref{eq:number8} and \ref{eq:number9}.
One can consider only contributions from the $u$ and $\bar{u}$ quarks.
A more complete analysis would also include small contributions from $d,\bar{d},s,\bar{s}$ and $c,\bar{c}$ pairs. 
With these approximations, the asymmetry is given by

\begin{equation}
A(x_1, x_2) \simeq
\frac{N^+_{u,G} + N^+_{G,u} 
        + N^+_{u, \bar{u}} + N^+_{\bar{u}, u}
	+N^+_{\bar{u},G} + N^+_{G,\bar{u}} }
{N^-_{u,G} + N^-_{G,u} 
        + N^-_{u, \bar{u}} + N^-_{\bar{u}, u}
	+N^-_{\bar{u},G} + N^-_{G,\bar{u}}} \, , 
\label{eq:number10}
\end{equation}
where $N^{\pm}_{i,j} = N^{\pm}_{i,j}(x_1, x_2)$, $i,j = u, \bar{u}$ and $G$. 
Introducing variables for the ratios of $\bar{u}$ and $u$ quark distribution 
functions $\bar{u}(x)$ and $u(x)$ to gluons distribution function $G(x)$, 
\[
r(x) = {{\bar{u}(x)} \over {G(x)}} \hskip 0.03 in {\mbox {,}} 
\hskip 0.3 in R(x) = {{u(x)} \over {G(x)}} 
\]
and for the ratio of the integrated Compton cross section to
the annihilation cross section, separately for the 
forward $(+)$ and backward $(-)$ angular regions,
\begin{equation}
B^{\pm} = {{\sigma^{\pm}_{u,G}} \over 
{\sigma^{\pm}_{u,\bar{u}}}}.
\label{eq:number12}
\end{equation}
Noting that $\sigma^{+}_{u,\bar{u}} =\sigma^{-}_{u,\bar{u}}$ and 
$\sigma^{+}_{u,G} = \sigma^{-}_{G,u} = 
\sigma^{+}_{\bar{u},G} = \sigma^{-}_{G,\bar{u}}$, 
the approximate expression for the asymmetry reduces to:
\begin{equation}
A(x_1, x_2; r(x_2)) = {{c_2 r(x_2) + c_1 r(x_1) + c_0} \over 
                       {d_2 r(x_2) + d_1 r(x_1) + d_0}} \, , 
\label{eq:number11}
\end{equation}
where the constants $c_i$ and $d_i$, $i = 2, 1,$ and $0$, defined in 
equation \ref{eq:number11}, are determined from the well known parton
cross sections and $u$ quark and gluon structure functions: 
\[
c_2 \sim R(x_1) + B^- 
\hskip 0.01 in {\mbox {,}} \hskip 0.3 in 
d_2 \sim R(x_1) + B^+ 
\]
\[
c_1 \sim R(x_2) + B^+ 
\hskip 0.01 in {\mbox {,}} \hskip 0.3 in 
d_1 \sim R(x_2) + B^- 
\]
\[
c_0 \sim B^+ R(x_1) + B^- R(x_2) 
\hskip 0.03 in {\mbox {,}} \hskip 0.3 in 
d_0 \sim B^- R(x_1) + B^+ R(x_2) 
\]

For full acceptance, with cross sections dominated by the 
very large $\cos(\vartheta_{\gamma}) = \pm 1$  upper limit of 
integration, the integrals defined in equations \ref{eq:number8} and 
\ref{eq:number9} yield
$B^+ = 3 / 8$ and $B^- = 0$. By restricting the upper limits of
integration to $\cos(\vartheta_{\gamma}) = \pm 0.8$, a more reasonable 
limit for a measurement at $RHIC$, $B^+$ increases to about
0.45 and $B^-$ increases to about 0.1. Using these higher values, it
is possible to estimate the size and range of observable asymmetries.

Choosing $x_1$ to be in the region where valence quarks dominate
the structure function ($x_1 \simeq 0.2$; $R(x_1) \simeq 0.96$ )  
and choosing $x_2$ in the region where the gluon distribution dominates 
($x_2 \simeq 0.05$; $R(x_2)\simeq 0.25$ ), the observable
range of asymmetry is $2.2 > A(x_1, x_2) > 0.75$. This corresponds to 
$r(x_2)$ ranging from zero to infinity while holding $r(x_1)$ at a 
fixed nominal value. For a nominal value of 
$r(x_2) = 0.088$,  $A(x_1, x_2)$ would be about 1.65. 

It should be a very general conclusion that the asymmetry will
be a ratio of linear functions of $r(x_1)$ and $r(x_2)$ as 
shown in equation \ref{eq:number11}.
The values of the constants $c_i$ and $d_i$ will be evaluated with
the $PYTHIA$ $Monte$ $Carlo$ program for various $(x_1, x_2)$ bins,
taking into account the experimental acceptance and $p_t$ cutoffs 
(see section \ref{sec:pythia}). It is noted that the
dependence of the asymmetry upon the detailed model of the acceptance 
appears to be modest. The ratio of asymmetries  for full acceptance 
($0.7 < |\cos(\vartheta_{\gamma})| < 1.0$)
to that for the limited acceptance described
above ($0.7 < |\cos(\vartheta_{\gamma})| < 0.8$) is only about 3:2, 
even as the formal cross sections range from infinite to finite.
The range of asymmetries for the cuts discussed in the beginning
of this section extends from about 0.75 to 2.2 reflecting
$\bar{u}(x_2)$ densities ranging from infinity to zero respectively.

It is an important feature of this analysis procedure that the
asymmetry is
relativitely insensitive to some
of the problems which have made the analysis of direct photon data difficult.
Accurate modeling of the detector for a
rapidly rising cross section near
the edge of acceptance in $\cos(\vartheta_{\gamma})$
has been one such problem. In this analysis, it is only the ratio
of forward to backward efficiencies that must be understood.
The price paid for this feature is that the inputs and outputs are not
the quark or antiquark structure functions directly but are the ratios of 
those structure functions to the gluon structure function.

With input from the ratios of quark to gluon structure functions
and the ratios of integrated parton Compton
cross sections to quark--antiquark annihilation cross sections,
the constants $c_i$ and $d_i$ are determined for various 
$(x_1, x_2)$ regions. These, along with the measured asymmetry 
$A(x_1, x_2)$, will result in a determination of the antiquark to glue ratio. 
More specifically, this will determine a well defined combination of
$r(x_2)$ and $r(x_1)$, with a rather small coefficient for $r(x_1)$:
\begin{equation}
r(x_2) + \varepsilon r(x_1) = r_{\varepsilon} \, , 
\label{eq:number13}
\end{equation}
where $\varepsilon$ and $r_{\varepsilon}$ are determined from 
the constants $c_i, d_i$ and the value of asymmetry $A(x_1, x_2)$: 
\[
\varepsilon = {{c_1 - d_1 A(x_1, x_2)} \over {d_2 A(x_1, x_2) - c_2}}
\hskip 0.05 in {\mbox {,}} \hskip 0.3 in 
r_{\varepsilon} = 
{{c_0 - d_0 A(x_1, x_2)} \over {d_2 A(x_1, x_2) - c_2}}.
\]

By considering different
regions $(x_1,x_2)$, each providing a linear equation in $r(x_1)$ and
$r(x_2)$, unambiguous analysis should be  possible.
Unless the ratio $r(x)$ grows rapidly with $x$ in the region
$x_2<x<x_1$, in contradiction of the 
conventional structure function sets, the effect of the 
$r(x_1)$ term is small, $\varepsilon \simeq 0.1$ (see section 
\ref{sec:pythia}). It is in this sense, that the asymmetry 
$A(x_1,x_2)$ is a rather direct measurement of
$r(x_2) = {\bar{u}(x_2) / G(x_2)}$. 

\newpage 

\section{PYTHIA simulation}
\label{sec:pythia}
In this section we present a discussion of the $p + p \rightarrow \gamma + jet$ 
process at $\sqrt{s} = 500 \, GeV$ based on the $PYTHIA$ simulation code 
\cite{sjostrand}. The kinematic cuts are $p_t > 10 \, GeV$, 
$-1 < \eta_{\gamma} < 2$ and $-1 < \eta_{jet} < 2$, where 
$\eta_{\gamma}$ and $\eta_{jet}$ are 
pseudo-rapidity of the direct photon and jet, respectively. 
The $p_t$  cut is high enough to provide a fairly clean 
measurement of a direct (hard) photon with a managable background 
from $\pi^0$ and $\eta$ meson decay. The $\eta_{\gamma}$ 
and $\eta_{jet}$ pseudo-rapidity cuts are consistent with the $STAR$ detector 
acceptances assuming a $Barrel$ and an $Endcap$ $Electro$--$magnetic$ 
$Calorimeter$ ($EMC$). 

With one $Endcap$ $EMC$, the $STAR$ detector acceptance in pseudo-rapidity 
has a natural asymmetry -- the position of the $Endcap$ $EMC$ in the 
pseudo--rapidity axis $1 < \eta < 2$. Thus, the symmetry between two proton 
beam directions is
broken and the side with the  $Endcap$ is defined to
be at  positive displacement from the interaction point and to have
acceptance for positive rapidity.
The convention will be that $x_1$ is defined to be a momentum fraction of
partons in the beam which moves in the positive direction,
consistent with equations \ref{eq:number3}--\ref{eq:number5}.

The cross section for the direct photon plus jet production in 
the kinematic region defined by these rapidity cuts are shown
for various $(x_1, x_2)$ bins in table \ref{tab:tab1}.
In table \ref{tab:tab2}  and table \ref{tab:tab3} the
cross sections from the two main subprocesses, pair annihilation and the 
Compton scattering, are shown. The total direct photon and jet
production cross section over this region is $9.3~ nb$.

\begin{figure}[tb]
        \begin{center}
        \leavevmode 
        \epsfxsize=0.8\hsize
        \epsfbox{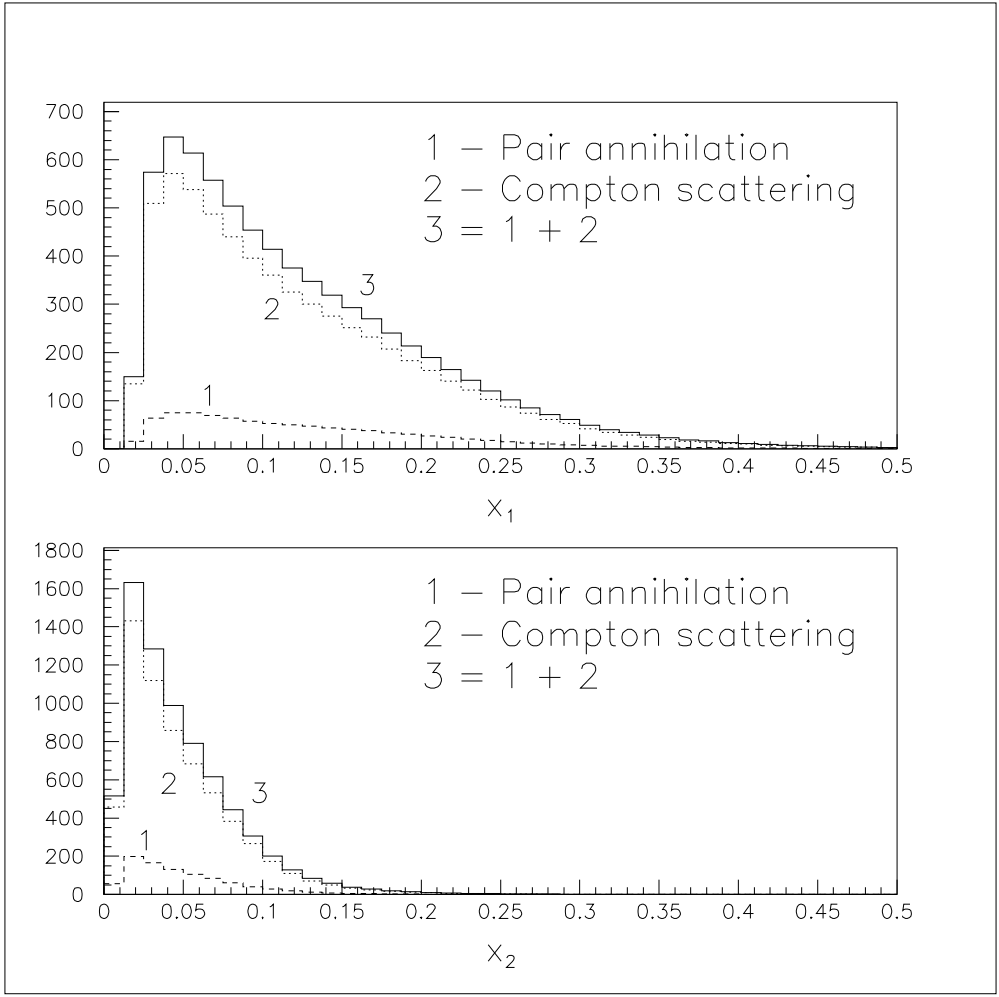}
\end{center}                         
\caption{Momentum fraction $x_1$ $(top)$ and $x_2$ $(bottom)$ distributions 
from the reaction $pp \rightarrow \gamma + jet$ at $pp$ center of mass 
energy $500~GeV$ and $p_t > 10 ~GeV$ in the acceptances of 
$-1 < \eta_{\gamma} < 2$ and $-1 < \eta_{jet} < 2$ for photon and 
jet respectively. The numbers in the vertical axes are cross sections in 
picobarms. Solid curves -- total contribution from pair annihilation and 
Compton scattering subprocesses, dashed curves -- contributions from the 
$(q\bar{q})$ pair annihilation subprocess, and dotted curves -- 
contributions from the Compton scattering subprocess.
}
\label{fig:fig1}
\end{figure}

Figure \ref{fig:fig1} shows the accepted cross section 
distribution of $x_1$ and $x_2$.
As shown in figure \ref{fig:fig1}, the pair annihilation 
contribution to the direct photon and jet production cross section is 
less than $10 \%$ (see, also, tables \ref{tab:tab2} and \ref{tab:tab3}). 
Two dimensional distributions of $x_1$ and $x_2$ are shown in figure 4. 
The asymmetry between $x_1$ and $x_2$ reflects the asymmetric 
pseudo--rapidity cuts: $-1 <  \eta_{jet} < 2$
and  $-1 < \eta_{\gamma} < 2$ ,
reflecting the $STAR$ $Barrel+Endcap$ coverage.

\begin{figure}[tb]
        \begin{center}
        \leavevmode 
        \epsfxsize=0.8\hsize
        \epsfbox{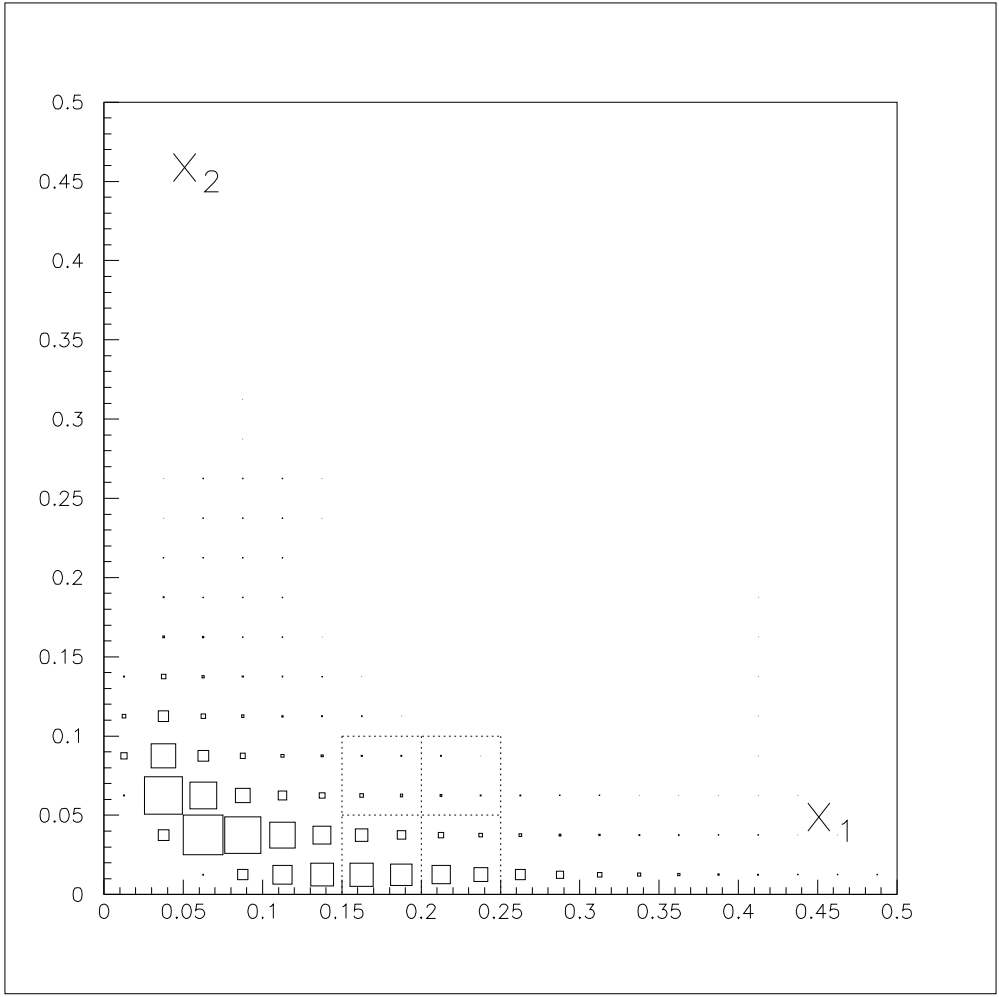}
\end{center}                         
\caption{Two dimentional $(x_1 \times x_2)$ plots for events distribution 
from the reaction $pp \rightarrow \gamma + jet$ at center of mass energy 
$500~GeV$ and $p_t > 10 ~GeV$. The acceptances of the photon and jet are 
$-1 < \eta_{\gamma} < 2$ and $-1 < \eta_{jet} < 2$, respectively. 
The events angular distributions from the four dotted squares 
with the centers in the points $(x_1, x_2) = (0.175, 0.075)$, 
$(0.225, 0.075)$, $(0.175, 0.025)$ and $(0.225, 0.025)$ are discussed in 
this paper.
}
\label{fig:fig2}
\end{figure}

The four small squares in figure \ref{fig:fig2} are the regions analyzed
in this paper. The values of constants $c_i$, $d_i$ normalized by $d_2 = 1$, 
$\varepsilon$ and values of asymmetry $A(x_1, x_2)$ from these four regions 
are presented in table \ref{tab:tab4}. Also presented in table \ref{tab:tab4} is the accuracy of the predicted asymmetry $A(x_1, x_2)$ from $PYTHIA$ 
$Monte$ $Carlo$ simulation code. The $\cos (\vartheta_{\gamma})$ 
distributions of the events from one of these four regions is presented 
in figure \ref{fig:fig3}. 

The minimum value of $x_2$,  consistent with the cuts mentioned above
in rapidity, $p_t$ 
and $\cos (\vartheta_{\gamma})$, is $x_2^{min} \simeq 0.02$.
The choice of a $\cos (\vartheta_{\gamma})$ limit value of $0.7$
is a trade--off between the need for a large asymmetry $A(x_1, x_2)$ 
and for a sufficiently large cross section.
Reduction of the $p_t$ limit would provide coverage in regions of 
$x_2 < 0.02$. Lower $x_2$ can also be reached by decreasing the
$\cos (\vartheta_{\gamma})$ limit. If asymmetry was defined by 
$|\cos (\vartheta_{\gamma})| > 0.5$, there is sensitivity down to 
$x_2^{min} \simeq 0.01$ with the same $p_t$ cut. 
Such a choice involves larger accepted cross sections
but weaker dependence of asymmetry upon the anti--quark ratio.

\begin{figure}[tb]
        \begin{center}
        \leavevmode 
        \epsfxsize=0.8\hsize
        \epsfbox{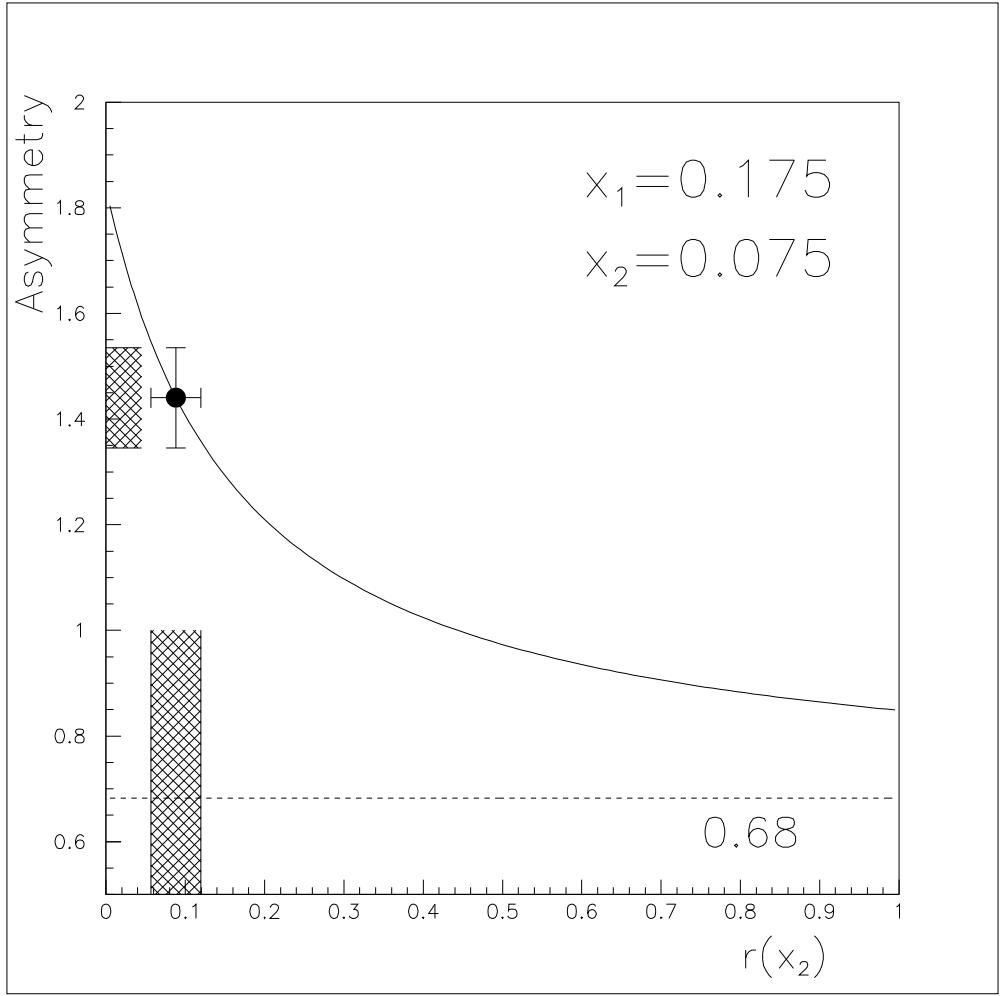}
\end{center}                         
\caption{The $r(x_2)$ dependences of the forward -- backward asymmetry 
$A(x_1,x_2;r(x_2))$ at $x_1 = 0.175$ and $x_2 = 0.075$ from the reaction 
$pp \rightarrow \gamma + jet$ at $pp$ center of mass energy $500~GeV$ and 
$p_t > 10 ~GeV$ in the acceptances of $-1 < \eta_{\gamma} < 2$ and 
$-1 < \eta_{jet} < 2$ for photon and jet respectively. 
The number under the dashed curve is the value of the 
forward--backward asymmetry at the limit $r(x_2) \gg 1$.
}
\label{fig:fig5}
\end{figure}

\begin{figure}[tb]
        \begin{center}
        \leavevmode 
        \epsfxsize=0.8\hsize
        \epsfbox{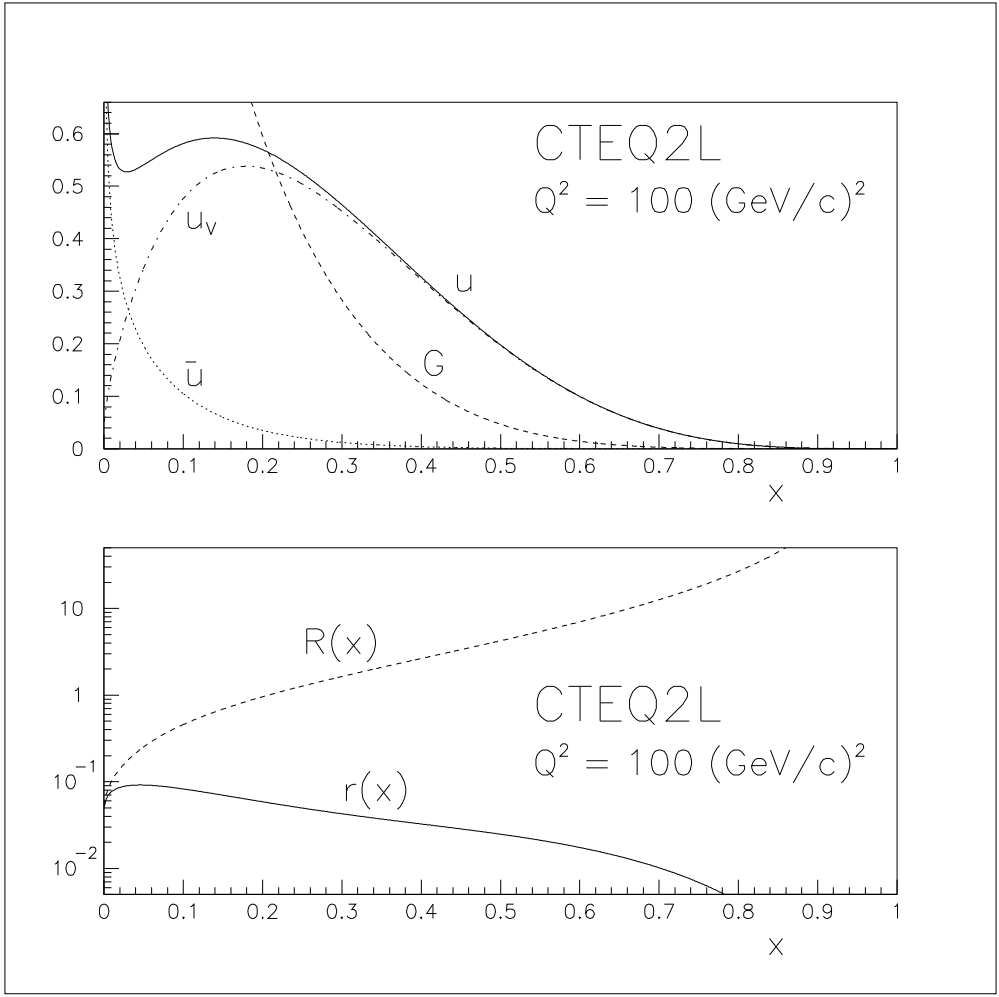}
\end{center}                         
\caption{The distributions of $u$, $\bar{u}$ quarks and $G$ gluons 
in the proton multiplied by $x$ $(top)$ and functions $r(x)$ and $R(x)$ at 
$Q^2 = 100 ~GeV^2$ from $CTEQ2L$ parametrization $(bottom)$. In the $top$ figure: 
solid curve -- $x u(x)$ quark distribution, dotted curve -- $x \bar{u}(x)$ 
antiquark distribution, dash--dotted curve -- $x u_v(x)$ valence quark 
distribution and dashed curve -- $x G(x)$ gluons distribution; the the $bottom$ 
figure: solid curve -- $r(x)$ and dashed curve -- $R(x)$. 
}
\label{fig:fig6}
\end{figure}

The dependence of the forward--backward asymmetry upon $r(x_2)$ is
presented in figure \ref{fig:fig5}. The point at $r(x_2) \simeq 0.088$ 
corresponds to the nominal $CTEQ2L$ structure function sets which were 
used in $PYTHIA$ (see figure \ref{fig:fig6}). For the four $(x_1, x_2)$ 
regions considered here, when the nominal $r(x_2) \simeq 0.088$ is 
assumed, the calculated asymmetries range from about 1.5 to 2.0.

\section{Analysis of Errors}
\label{sec:error}
\begin{figure}[tb]
        \begin{center}
        \leavevmode 
        \epsfxsize=0.8\hsize
        \epsfbox{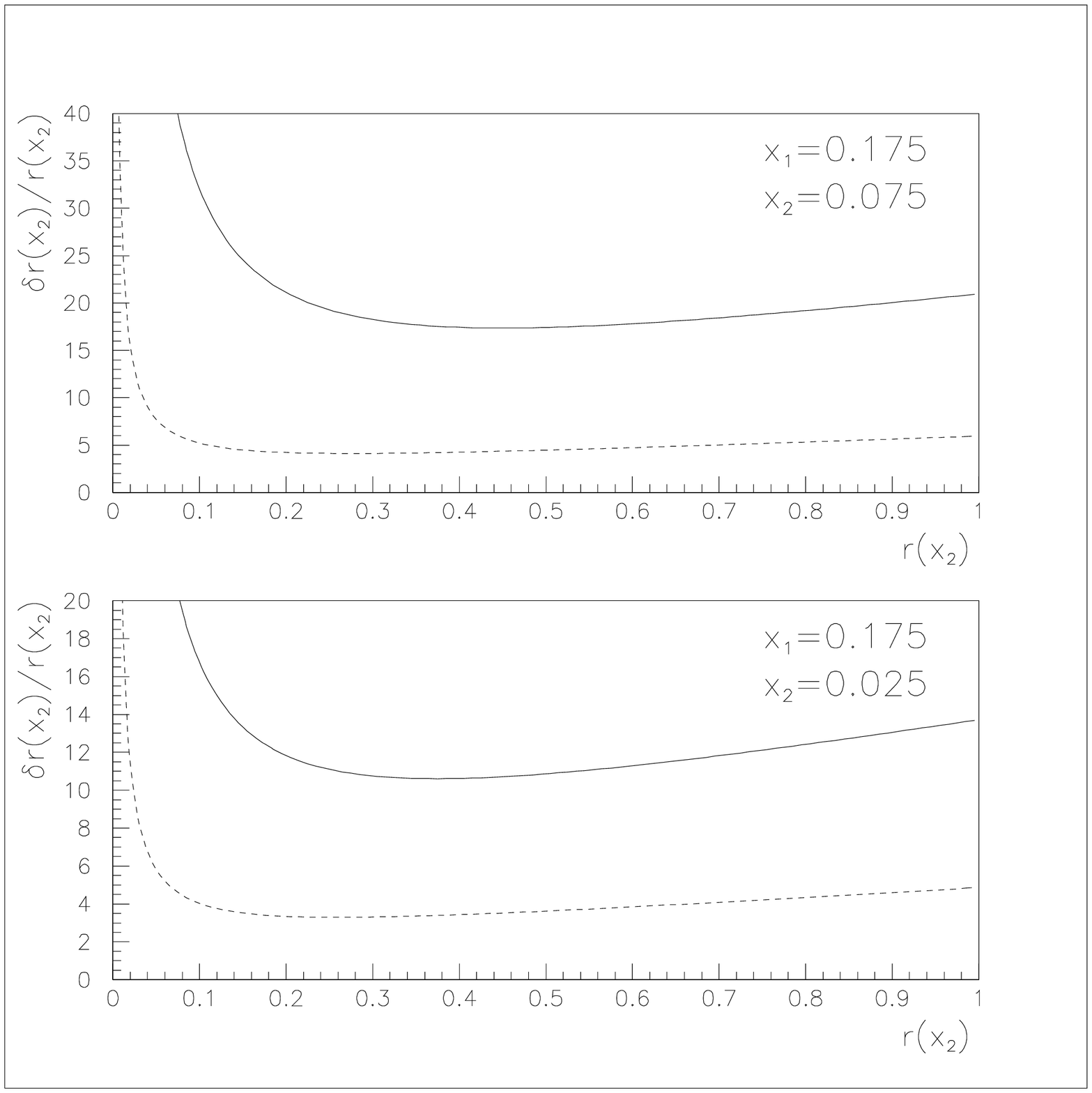}
\end{center}                         
\caption{The $r(x_2)$ dependences of the relative error 
$\delta r(x_2) / r(x_2)$ in percentages at different points of $x_1$ and 
$x_2$: 
$top$ -- $x_1 = 0.175, \, x_2 = 0.075$; 
$bottom$ -- $x_1 = 0.175, \, x_2 = 0.025$. 
Solid curves -- relative error $\delta r(x_2) / r(x_2)$ related with 
uncertainty of $R(x)$, if $\delta R(x) / R(x) = 10 \%$, dashed 
curves -- uncertainty on $r(x_2)$ related with statistical error from 
asymmetry. The value of the relative error of forward -- backward asymmetry 
$\delta A / A = 1 \%$, which would correspond to a standard $STAR$ run.
}
\label{fig:fig7}
\end{figure}

To compare event rates for a variety of experiments at
$RHIC$ involving proton--proton colliding beams, 
$RHIC$ experimentalists have agreed to consider a standard set
of integrated luminosities and energies. 
At center of mass energy   $\sqrt{s}=500~ GeV$, 
the standard luminosity of  $800 \, pb^{-1}$ is
used to evaluate the expected statistical error.  
The fractional statistical error, ${\delta A / A}$, for the
four regions considered is presented in table \ref{tab:tab4}.
The typical statistical error in measurement of ${\delta A / A}$
will be about 1\%. The corresponding error introduced in
the determination of ${\delta r(x_2) / r(x_2)}$ will be in the 
(3--7)\% range (see figure \ref{fig:fig7}), depending upon the 
actual value of $r(x_2)$. 

Systematic errors will be larger than statistical errors.
The systematic errors in the determination of $r(x_2)$ will come from various
sources. The important systematic error sources will involve uncertainties in
the determination of the constants in equations \ref{eq:number11} and 
\ref{eq:number13} and in the interpretation of the various sources of 
background. The following sources of errors have been considered. 

\begin{enumerate}
\item {\bf Contribution from partons other that $u$ quarks and gluons $G$.} 
        The effect on asymmetry from all the other quarks and antiquark 
        processes is small and calculable. Using the standard $CTEQ2L$, 
        the scale for the change in asymmetry $A(x_1, x_2)$ 
	due to the inclusion of other quarks and antiquarks is about
	5 \%. Even approximate simulation of these contributions would
	introduce a small uncertainty, ${\delta A / A}< 1 \%$. 
\item {\bf Uncertainty in determination of $c_i$ and $d_i$.} The inputs
	to this analysis are the ratio of up quark to gluon structure 
        functions $R(x)$. It will be assumed that the ratios can be 
        determined to 10\%.
	As seen in figure \ref{fig:fig7}, this implies an error in the
	(15--20) \% at $x_2 \simeq 0.025$ and (30--40) \% at 
        $x_2 \simeq 0.075$ range for $r(x_2)$. 
 	As mentioned above, the sensitivity to the detailed acceptance
	model near the limits of acceptance in rapidity and $p_t$ 
	is not great. 
\item {\bf Effects from beyond Leading Order calculations.} The main effect
	of the higher order corrections will be to increase the cross
	section by a $K$--factor. While this is a large increase in cross
	section, the asymmetry $A(x_1, x_2)$ is insensitive to the 
        normalization. A full higher order calculation would determine a 
        higher order correction to $A(x_1, x_2)$ and the analysis procedure 
        can be modified to take this into account. 

        It was seen that in a $Next$ $to$ $Leading$ $Order$ $(NLO)$ 
	calculation of the $CDF$ direct photon  cross section \cite{cdf1}, 
        the change in the $\cos (\vartheta_{\gamma})$ dependence is only at the 10\% level. This suggests that the sensitivity of 
        $A(x_1, x_2)$ to a full $NLO$ calculation may be rather modest.
\item {\bf Background from jet fragmentation to high--$p_t$ photons.}
	The standard background to direct photons is from unmatched
	photons from $\pi^0$ decays. Techniques have been developed to
	correct for this background \cite{cdf2}. It will be assumed here that
	the cross section of this background source can be directly 
	measured and subtracted. Techniques for this analysis are well
	developed and will not be discussed here
	as a source of error in this analysis.
	
	The background from photons produced directly in the jet fragmentation
	process is more difficult to estimate and subtract.
	The $CDF$ analysis of direct photon plus jet production in
	$p\bar{p}$ interactions shows that the angular dependence is
	relatively enhanced, in the forward--backward angle regions,
        over the $NLO$ prediction.
	The angular distribution is enhanced beyond that predicted in $NLO$
	by about $\sim (15 - 30) \%$ in the region 
	$\cos (\vartheta_{\gamma}) > 0.5$ \cite{cdf3}. 
        They attribute this discrepancy
	to an underestimate of the fragmentation function of jets into photons,
	a calculation done only in leading order. 
	This is about the level of modification  expected for  a full
	$NLO$ calculation of jet fragmentation.

	It is reasonable to assume, that the contamination
	of the direct photon signal from jet fragmentation will
	be symmetric in $\cos (\vartheta_{\gamma})$. This assumption can
	be checked by a full $NLO$ calculation of jet fragmentation into
	photons. For example, if the contamination
	was about 30\% and symmetric in $\cos (\vartheta_{\gamma})$, 
	$A(x_1, x_2)$ would be reduced from about $2.0$ to about $1.63$. 
        If the correction was known to be 
	$(30 \pm 10) \%$, the error would be ${\delta A / A} < 7 \%$. 
\item {\bf Effects from $k_t$.} A source of theoretical and experimental
	discrepancy is the initial state transverse momentum prior to a
	hard parton collision. An analysis of direct photon data \cite{kt} 
        suggests that this $k_t$ distribution is broader than expected,
	with a width as large as $(2 - 3) GeV/c$ in this kinematic region of 
        interest. Folded with the rapidly falling $p_t$ dependence, this 
        results in an excess cross section at the lower
	limit of $p_t$. In an analysis which anticipates a cross section
	proportional to the 
	structure function, this is a great complication.

	The crucial point here is that it is not $\bar{u}(x)$ which is measured
	in this analysis but the ratio $r(x) = \bar{u}(x) / G(x)$. This
	observable is likely to be a more slowly varying function of $x$ than
	the structure functions themselves in the small x region.
	For the analysis described here, a 50\% increase in the cross section
	due to these effects would not be much of a problem.
	The main effect would be to increase the uncertainty of $x_2$,
	somewhat smearing the bin over which the ratio $r(x_2)$ is 
        determined. The resulting error in $x_2$ is 
        ${\delta x_2 / x_2} \simeq (10 - 20) \%$. 
	We have shown that the asymmetry $A(x_1, x_2)$ varies rather slowly 
        with $x_2$ for the cases considered. It is noted in equation 
        \ref{eq:number3} that the ratio of $x_1 / x_2$ is well determined 
        by the rapidity and is not very sensitive to $k_t$ smearing. 
\end{enumerate}

It appears, that with reasonable expectations of an eventual  complete 
set of $NLO$ calculations, the largest source of error will come from 
the uncertainty in $R(x) = {u(x) / G(x)}$. It does not seem too optimistic 
to expect a determination of $r(x) = {\bar{u}(x) / G(x)}$ to an accuracy of 
$(15 - 30) \%$, at several values of $x$ in the 0.02 to 0.1 region,
using this method of analysis. The uncertainty in this measurement of 
the $\bar{u}$ structure function will be limited by
the uncertainty in the $u$ quark and gluon distributions. 

\newpage 

\section{Spin Dependence}
\label{sec:lspin}
The greatest interest in $pp$ measurements at $RHIC$ involves
measurements with polarized beams. With longitudinally polarized beams,
the dependence of the cross section on beam polarization can be measured
to determine the polarization of the quarks and gluons within the proton.
Another asymmetry, $A_{LL}$,  is defined to be
the difference between the cross section 
for spin aligned and anti-aligned protons.
This quantity is sensitive to the polarized parton distributions. 
A region is considered which is centered about $(x_1,x_2)$,
with an unpolarized
cross section $\sigma(x_1,x_2)$ given by the average of two 
polarized cross sections
$\sigma^{\uparrow \uparrow }(x_1,x_2)$ and 
$\sigma^{\uparrow \downarrow }(x_1,x_2)$, corresponding to longitudinal 
spins aligned or anti--aligned respectively.
$A_{LL}(x_1,x_2)$ is:
\begin{equation}
\begin{array}{ll}
A_{LL}(x_1,x_2)&=
\frac{
\sigma^{\uparrow \uparrow}(x_1,x_2)-\sigma^{\uparrow \downarrow}(x_1,x_2)
}
{
\sigma^{\uparrow \uparrow}(x_1,x_2)+\sigma^{\uparrow \downarrow}(x_1,x_2)
}\\
&~\\
&=\sum_{i,j}P^L_i(x_1) P^L_j(x_2) f_{i,j}(x_1,x_2)
 a^{LL}_{i,j}(\cos (\vartheta_{\gamma}))\\
\end{array}
\label{eq:Allsum}
\end{equation}
where $f_{i,j}(x_1, x_2)$ is the ratio of the unpolarized cross section 
from partons  $i$ and $j$ to the sum of all processes in the region 
$(x_1, x_2)$, $P^L_i(x)$ is the longitudinal polarization of a parton of type
$i$ at momentum fraction $x$ relative to the longitudinal spin of 
the proton, and $a^{LL}_{i,j}$ is the parton level analyzing power for 
this subprocess which can depend on the parton center of mass scattering angle.
For annihilation and Compton scattering, the analyzing power is \cite{spin}: 
\begin{equation}
a^{LL}_{q \bar{q}} = a^{LL}_{\bar{q} q} = -1 
\end{equation}
and
\begin{equation}
a^{LL}_{qG} = a^{LL}_{\bar{q}G} = {\frac {\hat{s}^2 - \hat{u}^2} {\hat{s}^2 + \hat{u}^2}} = 
            {{4 - (1 - \cos (\vartheta_{\gamma}))^2} \over 
             {4 + (1 - \cos (\vartheta_{\gamma}))^2}} 
\label{eq:parton:all}
\end{equation}
\begin{equation}
a^{LL}_{G q} = a^{LL}_{G \bar{q}} = {\frac {\hat{s}^2 - \hat{t}^2} {\hat{s}^2 + \hat{t}^2}} = 
            {{4 - (1 + \cos (\vartheta_{\gamma}))^2} \over 
             {4 + (1 + \cos (\vartheta_{\gamma}))^2}} 
\label{eq:parton:all_new}
\end{equation}

If the magnitude of the anti--up quark structure function
is near the nominal value
$r(x_2) \simeq 0.1$, then it is interesting to consider $A_{LL}^+$ and
$A_{LL}^-$ defined from considering cases of 
$\cos (\vartheta_{\gamma}) > 0.7$ and 
$\cos (\vartheta_{\gamma}) < -0.7$ respectively. 
Considering only the two most important terms in the sum from
equation \ref{eq:Allsum},
\begin{equation}
\begin{array}{l l}
A_{LL}(x_1,x_2) \simeq&
P_u(x_1)P_G(x_2)f_{u,G}(x_1,x_2) a^{LL}_{u,G}(\cos (\vartheta_{\gamma})) \\
&~\\
&
+ P_u(x_1)P_{\bar{u}}(x_2)f_{u,\bar{u}}(x_1,x_2) (-1) \\
\end{array}
\label{eq:twoterms}
\end{equation}
Many of the factors in equation \ref{eq:twoterms} are to some degree known.
In the region of $x_1$ considered here, $P_u(x_1) \simeq 0.3$ \cite{slac,emc}. 
The functions $f_{i,j}(x_1, x_2)$ are the fractional contribution from 
a process involving $i$ and $j$ in the initial state and can be evaluated
in $PYTHIA$. 

The result is that in the forward region, 
the gluon polarization which is also measured in other kinematic
regions will be nearly proportional to $A_{LL}^+$.
However, in the backward photon region, 
$A_{LL}^-$ will be as sensitive to the
anti-up quark polarization as to the gluon polarization. 
Including a degradation of the signal
from finite $(70 \%)$ beam polarization,
a $0.5 \%$ measurement of $A_{LL}^-$ would be required to determine
$P_{\bar{u}}(x_2)$ with an error of about 
$\delta P_{\bar{u}}(x_2) \simeq \pm 0.1$.

It is interesting to consider the complimentary spin asymmetry with
transversely polarized beams, $A_{NN}$. The formulas differ from
that of the longitudinal asymmetry in two ways. First, 
the observed transverse asymmetry is also proportional to 
$\sin (\varphi)$, with $\varphi$ the azimuthal angle of the production 
plane relative to the transverse polarization axis. 
Second, there is no contribution to the transverse
asymmetry from processes with gluons, therefore the leading source of
transverse asymmetry should be the anti--up quark annihilation process,
which will have a parton level analyzing power of $-1$. 
In the case of transverse polarization, none of the quark polarizations 
are known so detailed predictions are not very constrained, however,
the sensitivity to polarization could be similar to what is described
above for longitudinal $\bar{u}$ polarization in this region.

\section{Acknowledgments}
\label{sec:thanks}
This work has been supported in part by grant 
$RFFI$ $\#$ 95--02--05061 
from the Russian Academy of Science and the National Science Science Foundation.


\newpage

\begin{table}
\begintable%
$\Delta x_1$ | $.00 - .05$ | $.05 - .10$ | $.10  - .15$ | $.15 - .20$ | 
$.20 - .25$ | $.25 - .30$ | $.30 - .35$ | $.35 - .40$ \crnorule
$\Delta x_2$ | | | | | | | | \crthick
$.00 - .05$ | 426 | 1,640  | 1,460 | 1,060 | 660 | 355 | 
173 | 85 \cr
$.05 - .10$ | 1,260 | 812 | 264 | 112 | 51 | 26 | 13 | 
7 \cr
$.10 - .15$ | 365 | 154 | 45 | 20.0 | 9.0 | 5.2 | 2.9 | 1.8
\cr
$.15 - .20$ | 70 | 42 | 13.0 | 4.8 | 2.5 | 1.5 | 0.64 | 0.49 
\endtable

\caption{Pythia cross sections in picobarns of the direct photon and jet 
production
process. The  statistical errors for the entries in these tables correspond  
to 35 events per picobarn.
}
\label{tab:tab1}
\addtocontents{lot}{\protect\addvspace{.10in}}
\end{table}
\begin{table}
\begintable%
$\Delta x_1$ | $.00 - .05$ | $.05 - .10$ | $.10  - .15$ | $.15 - .20$ | 
$.20 - .25$ | $.25 - .30$ | $.30 - .35$ | $.35 - .40$ \crnorule
$\Delta x_2$ | | | | | | | | \crthick
$.00 - .05$ | 26 | 158 | 162 | 128 | 80 | 44 | 22 | 
11 \cr
$.05 - .10$ | 124 | 116 | 44 | 20 | 8.4 | 4.7 | 2.7 | 1.2 \cr
$.10 - .15$ | 37.64 | 23.07 | 7.36 | 3.26 | 1.89 | 1.25 | 0.73 | 0.20 \cr
$.15 - .20$ | 7.04 | 6.46 | 1.92 | 0.73 | 0.26 | 0.23 | 0.20 | 0.15 
\endtable
\caption{Cross sections in picobarns from the annihilation subprocess.}
\label{tab:tab2}
\addtocontents{lot}{\protect\addvspace{.10in}}
\end{table}
\begin{table}
\begintable%
$\Delta x_1$ | $.00 - .05$ | $.05 - .10$ | $.10  - .15$ | $.15 - .20$ | 
$.20 - .25$ | $.25 - .30$ | $.30 - .35$ | $.35 - .40$ \crnorule
$\Delta x_2$ | | | | | | | | \crthick

$.00 - .05$ | 400 | 1,480 | 1,290 | 935 | 582 | 311 | 
152 | 74 \cr
$.05 - .10$ | 1,140 | 697 | 220 | 91 | 42 | 21 | 10 | 
6 \cr
$.10 - .15$ | 328 | 130. | 37 | 16 | 7.1 | 4. | 2.0 | 1.6 \cr
$.15 - .20$ | 63 | 35 | 11 | 4.1 | 2.2 | 1.22 | 0.64 | 0.35 
\endtable
\caption{Cross sections in picobarns from the Compton scattering subprocess.}
\label{tab:tab3}
\addtocontents{lot}{\protect\addvspace{.10in}}
\end{table}
\begin{table}
\begintable%
$x_1$ and $ x_2$ | $\sigma^{+}$ | $\sigma^{-}$ | $\varepsilon$ | 
$c_2$ | $c_1$ | $c_0$ | $d_1$ | $d_0$ | $A(x_1, x_2)$ | 
$\delta A / A$ $(\%)$ | $\delta A / A$ $(\%)$ \crnorule 
| (pb) | (pb) | | | | | | | | $PYTHIA$ | $STAR$ run \crthick 
$0.175, 0.075$ | 18.7 | 13.0 | 0.13 | 0.68 | 0.59 | 0.27 | 
0.34 | 0.15 | 1.44 | 3.8 | 1.3 \cr
$0.225, 0.075$ | 9.0 | 5.6 | 0.03 | 0.75 | 0.49 | 0.29 | 
0.28 | 0.13 | 1.61 | 5.5 | 1.9 \cr
$0.175, 0.025$ | 33.9 | 19.0 | 0.12 | 0.70 | 0.49 | 0.31 | 
0.20 | 0.13 | 1.78 | 2.9 | 1.0 \cr
$0.225, 0.025$ | 10.4 | 5.5 | 0.06 | 0.84 | 0.44 | 0.33 | 
0.19 | 0.13 | 1.87 | 5.4 | 1.9 
\endtable
\caption{Cross sections and prediced asymmetry $A(x_1, x_2)$ parameters as
defined in Equation \ref{eq:number11}. The parameters are calculated with
subprocess cross sections from Pythia and with $d_2 \equiv 1$.} 
\label{tab:tab4}
\addtocontents{lot}{\protect\addvspace{.10in}}
\end{table}

\newpage

\newpage

\begin{references}
\bibitem{h1} {\it H1 Collab.}, S.~Aid {\it et al.}, Nucl. Phys. 
              {\bf D470}, 3 (1996).
\bibitem{zeus} {\it ZEUS Collab.}, M.~Derrick {\it et al.}, Z. Phys. 
                {\bf C69}, 607 (1996); 
                preprint DESY 96 - 076 (1996), Z. Phys., to be published. 
\bibitem{duke} D.~W.~Duke and J.~F.~Owens, Phys. Rev. {\bf D30},
                49 (1984).
\bibitem{cteq95} H.~L.~Lai {\it et al.}, Phys. Rev. {\bf D51}, 4763 (1995).
\bibitem{cteq93} J.~Botts {\it et al.}, Phys. Lett. {\bf B304}, 159 (1993).
\bibitem{morfin} J.~G.~Morfin and W.~K.~Tung, Z. Phys. {\bf C52}, 13 (1991).
\bibitem{martin} A.~D.~Martin, R.~G.~Roberts and W.~J.~Stirling, 
                   Phys. Lett. {\bf B387}, 419 (1996); 
                   {\it ibid.} {\bf B356}, 89 (1995); 
                   {\it ibid.} {\bf B354}, 155 (1995); 
                   Phys. Rev. {\bf D50}, 6734 (1994). 
\bibitem{halzen} F.~Halzen and D.~M.~Scott, Phys. Rev. {\bf D18},
                  3378 (1978).
\bibitem{owens} J.~F.~Owens, Rev. Mod. Phys. {\bf 59}, 465 (1987).
\bibitem{sjostrand} T.~Sj\"{o}strand, Comput. Phys. Commun. {\bf 82}, 
                      74 (1994).
\bibitem{cdf1} {\it CDF Collab.}, F.~Abe {\it et al.}, Phys. Rev. Lett. 
                {\bf 71}, 679 (1993). 
\bibitem{cdf2} {\it CDF Collab.}, F.~Abe {\it et al.}, Phys. Rev. Lett. 
                {\bf 73}, 2662 (1996); {\it ibid.} {\bf 68}, 2734 (1992). 
\bibitem{cdf3} {\it CDF Collab.}, L.~Nodulman, {\it 28th International 
        Conference on High Energy Physics (ICHEP '96)}, Warsaw, Poland, 
        July 1996; FERMILAB - Conf - 96/337 - E (1996). 
\bibitem{kt} J.~Huston {\it et al.}, preprint MSU - HEP - 41027, 
               CTEQ - 407 (1995). 
\bibitem{spin} C.~Bourrely {\it et al.}, Phys. Rep. {\bf 177}, 319 (1989).
\bibitem{slac} {\it SLAC E - 80}, M.~J.~Alguard {\it et al.}, Phys. Rev. Lett. 
               {\bf 37}, 1261 (1976); {\it ibid.} {\bf 41}, 70 (1978). \\ 
               {\it SLAC E - 130}, G.~Baum {\it et al.}, Phys. Rev. Lett. 
               {\bf 51}, 1135 (1983). 
\bibitem{emc} {\it EMC}, J.~Ashman {\it et al.}, Phys. Lett. {\bf B206}, 
              364 (1988); Nucl. Phys. {\bf B328}, 1 (1989). 
\end{references}
\end{document}